\documentstyle[hip-artc]{article}  
\volnumber{10}  \edyear{1999}  \frompage{333} \topage{343}                
\recrevdate{5 August  1999}                                              
\def\rightmark{Discontinuities in Relativistic Hydrodynamics} \def\leftmark{K. A. Bugaev et al.}
\def\ro{{\rm o}}
\def\fos{freeze-out shock\,}

\title{ \bf DISCONTINUITIES IN RELATIVISTIC\\ 
\vspace*{2mm}
HYDRODYNAMICS ACROSS SPACE-LIKE\\ 
\vspace*{4.4mm}
AND TIME-LIKE HYPERSURFACES
}

\authors{
{\twerm K. A. Bugaev$^{1,2}$, M. I. Gorenstein$^{1,2}$
and W. Greiner$^{1}$
}\\[2.812mm]
{\normalsize
\hspace*{-8pt}$^1$ Institut f\"ur Theoretische Physik, J.W. Goethe Universit\"at, \\ 
D-60054 Frankfurt, Germany\\[0.2ex] 
\hspace*{-8pt}$^2$ Bogolyubov Institute for Theoretical Physics,\\ 
252130 Kiev, Ukraine
}}
 
\abstract{
Discontinuities in relativistic hydrodynamics --
shock waves and freeze-out shocks --
are considered across both space-like and time-like hypersurfaces.
We analyze the peculiar features of the
freeze-out discontinuities and their connection to
the shock-wave phenomena.
}
 
\begin{document}
 
\maketitle

\section{Hydrodynamical Discontinuities}
 
Relativistic hydrodynamical model \cite{bib1} has
been widely discussed in recent years within their connection to
high energy nucleus-nucleus (A+A) collisions (see, for example,
\cite{bib2}).
The system evolution in relativistic hydrodynamics is governed by
the energy-momentum tensor
\begin{equation}\label{tmunu}
 T^{\mu\nu}~ =~ (\varepsilon + p)u^{\mu}u^{\nu}~ -~ pg^{\mu\nu}
\end{equation}
and conserved charge currents.
The baryonic current,
\begin{equation}\label{barcur}
 j^{\mu}~ =~ n~u^{\mu}~,
\end{equation}
plays the main role in the application to A+A collisions.

The hydrodynamical description includes
the local thermodynamical variables
(energy density $\varepsilon$, pressure $p$,
baryonic density $n$) and the collective
four-velocity $u^{\mu}=(1-{\bf v}^2)^{-1/2}(1, {\bf v})$.
The continuous flows are the solutions of the
hydrodynamical equations
\begin{equation}\label{hydroeq}
\partial _{\mu}T^{\mu\nu}~ =~0~,~~~~~
\partial _{\mu}j^{\mu}~ =~0~,
\end{equation}
with specified initial conditions.

Equations (\ref{hydroeq})
are the differential form
of the energy-momentum and baryonic number conservation laws. Along
with these continuous flows, the conservation laws can also be
realized in the form of discontinuous hydrodynamical flows which are
called shock waves and satisfy the following equations:
\begin{equation}\label{enmombar}
 T_\ro^{\mu\nu}\Lambda_{\nu} = T^{\mu\nu}\Lambda_{\nu}  \;\; ,~~~~
 n_{\ro}u_{\ro}^{\mu}\Lambda_{\mu}= n u^{\mu}\Lambda_{\mu} \;\; ,
\end{equation}
where $\Lambda _{\mu}$ is the unit 4-vector normal to the
discontinuity hypersurface.
In Eq.~(\ref{enmombar})
the zero subscript corresponds
to the initial state ahead of the shock front and
the quantities without
an index are the final state values behind it.
To complete the system of the
hydrodynamical equations (\ref{hydroeq})
or (\ref{enmombar})
the fluid equation of state (EoS) has to be added as an input
$p=p(\varepsilon ,n)$.

Besides the fluid EoS and the initial conditions
for the fluid evolution
the third crucial ingredient of the hydrodynamical model
of A+A collisions
is the so-called freeze-out (FO) procedure, i.e. the prescription
for the calculations of the final hadron observables.
Particles which leave
the system and reach the detectors are considered
via FO scheme,
where the frozen-out particles are formed
on a 3-dimensional hypersurface in space-time.
The FO
description has a straight influence on particle momentum spectra
and therefore on all hadron observables.
A generalization of the well known Cooper-Frye (CF)
formula \cite{bib3} to the case of time-like (t.l.)
FO hypersurface was suggested in Ref.~\cite{bib4}.
The new formula does not contain negative particle number
contributions on t.l. FO hypersurface appeared
in the CF procedure from those particles which cannot leave
the fluid during its hydrodynamical expansion.
The FO procedure of Ref.~\cite{bib4} has been further developed
in a series of publications [5-10].
The particle emission from the t.l. parts of the FO hypersurface looks
as a `discontinuity' in the hydrodynamic motion. We call it the FO shock.

In  the present paper we analyze the
discontinuities
in relativistic hydrodynamics -- normal shock waves and
the FO shocks -- across both space-like (s.l.) and t.l.
hypersurfaces.

\section{Relativistic Shock Waves}  
The relativistic shock waves are defined by
equations (\ref{enmombar}).
The important constraint on the shock transitions
(\ref{enmombar})
is the requirement of non-decreasing
entropy
(thermodynamical
stability condition):
 \begin{equation}\label{entropy}
su^{\mu}\Lambda _{\mu} \ge s_{\ro}u_{\ro}^{\mu}\Lambda _{\mu} \; ,
\end{equation}
where $s$ is the entropy density.

We consider one-dimensional hydrodynamical motion in what
follows. In its usual sense
the theory of the shock waves corresponds to
the discontinuities across the t.l. surface, i.e. the normal
vector $\Lambda _{\mu}$ is a s.l. one.
It means that the shock-front velocity
is smaller than 1.
In this case one can always choose
the Lorentz frame where the shock front is at rest. Then the
hypersurface of shock discontinuity is
$x_{sh} = const$ and $\Lambda _{\mu} =(0,1)$. The shock
equations (\ref{enmombar})
become
\begin{equation}\label{slshock}
T_{\ro}^{01} = T^{01}, \; \; T_{\ro}^{11} = T^{11} \;\; ,~~~
n_{\ro}u_{\ro}^{1} = nu^{1} \; \; .
\end{equation}
From Eq. (\ref{slshock}) one obtains
\begin{equation}\label{slvel}
v_{\ro}^{2} = \frac{(p - p_{\ro})(\varepsilon + p_{\ro})}
{(\varepsilon - \varepsilon_{\ro})(\varepsilon_{\ro} + p)} \; , \; \;
v^{2} = \frac{(p - p_{\ro})(\varepsilon_{\ro} + p)}
{(\varepsilon - \varepsilon_{\ro})(\varepsilon + p_{\ro})} \;\; .
\end{equation}
Substituting (\ref{slvel}) into the last equation in (\ref{slshock})
we obtain the well known
Taub adiabate (TA) equation ~\cite{bib11}
\begin{equation}\label{TA}
n^{2}X^{2} - n_{\ro}^{2}X_{\ro}^{2} - (p - p_{\ro})
(X + X_{\ro}) = 0 \;\; ,
\end{equation}
in which $X \equiv (\varepsilon + p)/n^{2}$, and TA therefore
contains only the thermodynamical variables.

The point $(\varepsilon_{\ro}, p_{\ro},n_{\ro})$ is called the center of
the TA.
The mechanical stability of the
shock transition from the state $(\varepsilon_{\ro}, p_{\ro},
n_{\ro})$ to $(\varepsilon, p, n)$
was studied in
Refs.~\cite{bib12,bib13}.
Thermodynamical stability (\ref{entropy}) follows from the mechanical
stability, and the inverse statement
is not in general true \cite{bib12,bib13}.
The consequences of the mechanical stability
are also the well-known inequalities for
the speed of sound
and the flow velocities at both sides of the shock
front in its rest frame
\begin{equation}\label{sound}
c_{s\ro}~\leq ~v_{\ro}~,~~~~c_{s}~\geq ~v~.
\end{equation}

In the thermodynamically normal media
the compression shocks are stable
whereas the rarefaction shocks become stable in the thermodynamically
anomalous media.
The shock-wave stability
in the case of the
phase transitions between hadron matter and the quark-gluon plasma
are discussed in Refs.~\cite{bib14,bib15}.

Let us consider now the
discontinuities on a hypersurface with a t.l. normal
vector $\Lambda _{\mu}$ (t.l. shocks). This new possibility
was suggested by Csernai in Ref.~\cite{bib16}.
In this case
one can always choose another convenient Lorentz
frame ('simultaneous system') where the hypersurface of the
discontinuity is
$t_{sh} = const$ and $\Lambda _{\nu}
= (1,0)$. Equations (\ref{enmombar}) become then
\begin{equation}\label{tlshock}
T_{\ro}^{00} = T^{00}, \; \; T_{\ro}^{10} = T^{10} \;\; ,~~~
n_{\ro}u_{\ro}^{0} = nu^{0} \;\; .
\end{equation}
From Eq. (\ref{tlshock}) we find
\begin{equation}\label{tlvel}
\tilde{v}_{\ro}^{2} =
\frac{(\varepsilon - \varepsilon_{\ro})(\varepsilon_{\ro} + p)}
{(p - p_{\ro})(\varepsilon + p_{\ro})}~, \; \; ~~
\tilde{v}^{2} = \frac
{(\varepsilon - \varepsilon_{\ro})(\varepsilon + p_{\ro})}
{(p - p_{\ro})(\varepsilon_{\ro} + p)} \;\;,
\end{equation}
where we use $``\sim"$ sign to distinguish the t.l. shock case
(\ref{tlvel}) from the standard s.l. shocks (\ref{slvel}).
Substituting (\ref{tlvel})
into the last equation in (\ref{tlshock})
one finds the equation for t.l. shocks  which is identical to
the TA  of Eq.~(\ref{TA}). We stress, however, that the intermediate steps
are
quite different. The two solutions,
Eqs.~(\ref{tlvel})
and (\ref{slvel}),
are connected to each other by simple relations,
\begin{equation}\label{tlslvel}
\tilde{v}_{\ro}^{2} =
\frac{1}{v_{\ro}^{2}}~, \; \;~~
\tilde{v}^{2} = \frac
{1}{v^{2}} \;\; ,
\end{equation}
between the velocities of s.l. and t.l. shocks. These relations
show that only one type of transitions can be realized for the given
initial  and final states.
Physical regions $[0,1)$ for $v_{\ro}^{2},v^{2}$ (\ref{slvel}) and for
$\tilde{v}_{\ro}^{2},\tilde{v}^{2}$ (\ref{tlvel}) in
the $(\varepsilon $--$p)$-plane are shown
in \mbox{Fig. 1.}

When the initial and final states
are thermodynamically equilibrated
the TA passes through the point $(\varepsilon _{\ro},p_{\ro})$
and lies as a whole in the regions
I and IV in {\mbox Fig.~ 1} \cite{bib13}, i.e. only compression
and/or rarefaction s.l. shocks
(with s.l. normal vector $\Lambda_{\mu}$) are permitted.

The only way to make the t.l. shocks to be possible
is to allow the metastable initial and/or final states.
Then new possibilities of t.l. shock
transitions (\ref{tlshock},\ref{tlvel}) to regions III and VI in Fig. 1
would be realized (see, e.g.,  the t.l. shock hadronization of
the supecooled
quark-gluon plasma  in Ref.~\cite{bib17}).

\begin{figure}[htb]
\vspace*{-0.5cm}
                 \insertplot{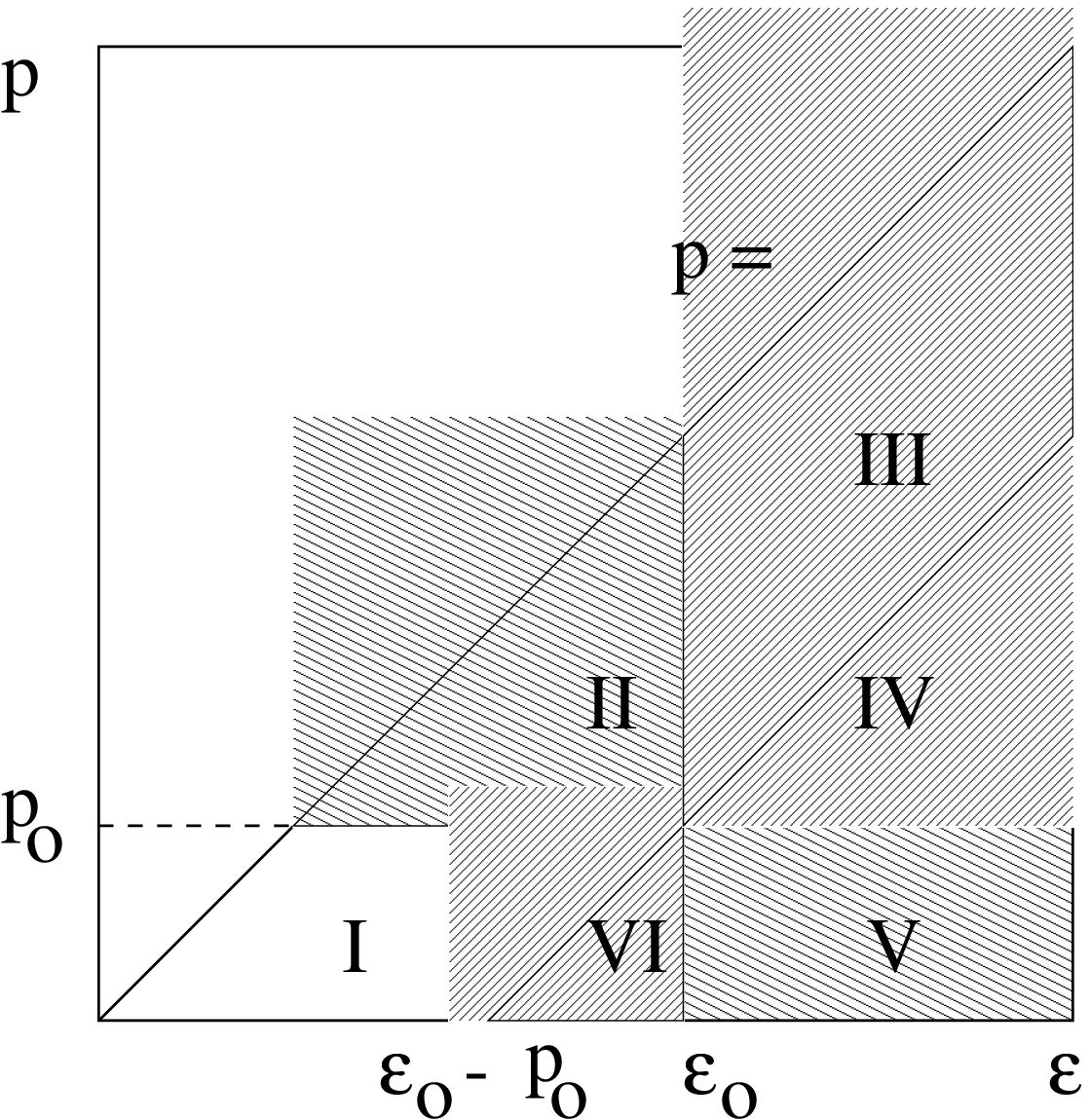}
\vspace*{-0.9cm}
\caption[]{
Possible final states in the $\varepsilon -p$ plane for
shock transitions from the initial state $(\varepsilon_{\ro} ,p_{\ro}$).
I and IV are the physical regions for s.l. shocks, III and VI
for t.l. shocks. II and V are unphysical regions for both
types of shocks. Note, that only states with $p\le \varepsilon$ are
possible for any physical equation of state in relativistic theory.
}
\label{fig1}
\end{figure}

\section{Freeze Out Shocks}

In the hydrodynamical model of A+A collisions the fluid expansion
has to be ended by the correct transformation
of the fluid into free streaming
hadrons. The most important requirement for this FO procedure
is to satisfy energy-momentum and charge conservations
during the fluid transition into free final particles.
It is assumed that there is a narrow space-time region where
the mean free path of the fluid constituents increases
rapidly and becomes comparable with the characteristic
size of the system. The local thermal equilibrium is supposedly
maintained in this intermediate region. In practice, one considers
a ``zero width'' approximation and introduces a FO surface,
so that particle distributions remain frozen-out from there on.

In his original paper Landau \cite{bib1} defined the
FO hypersurface by the condition \mbox{$T(t,x)=T^*$}, where
$t$ and $x$ are the time and the space coordinate respectively, and $T^*$
is the fluid FO temperature chosen to be approximately equal to the
pion mass.
The FO procedure, first introduced by Milekhin \cite{bib19}, was
improved further by Cooper and Frye \cite{bib3} and
this method has been used ever since.
Final momentum spectra for {\it i}-th type hadrons are
expressed by the formula
\cite{bib3}:
\begin{equation}\label{CF}
k^{\ro} \frac{d^3N_i}{dk^3}~=
~\frac{d_i}{(2\pi)^3}\int_{\Sigma_f}d\sigma_{\mu}k^{\mu}~
\phi_i\left(\frac{k^{\nu}u_{\nu} - \mu_i}{T^*}\right)~,
\end{equation}
where $d\sigma_mu$ is the external normal 4-vector
to the FO hypersurface $\Sigma_f$. It equals to
$ S_{tr}(-dx,dt)$ in 1+1 dimension, where $S_{tr}$
is the (constant) transverse area of the system.
$\phi_i$ denotes the local thermal
Bose or Fermi distributions, $d_i$ is the degeneracy factor for
particle $i$ with the chemical potential $\mu_i$.
The particle spectrum (\ref{CF}) is known as the CF
distribution function.

The initial conditions of the hydrodynamical motion
are given on the initial (s.l.) hypersurface $\Sigma_{in}$.
The final hypersurface
$\Sigma_{f}$ should be closed to $\Sigma_{in}$ and
in general $\Sigma_{f}$ consists of both s.l.
and t.l. parts.
The CF formula(\ref{CF}) leads automatically to
the energy-momentum and charge number conservations,
if the ideal gas fluid EoS at the FO hypersurface
is assumed.

The CF formula (\ref{CF}) still does
not provide a complete solution of the FO problem.
The FO surface consists of t.l. parts too.
Eq.~(\ref{CF}) can not however be used for a t.l. FO surface
(i.e., s.l. normal vector $d\sigma_{\mu}$):
free final particles ``return'' to the fluid if $d\sigma_{\mu}k^{\mu} < 0$,
and this causes unphysical {\it negative} contributions to the
{\it number} of final particles.
The modified
FO procedure and new formula for the final particle spectra
emitted from the t.l. FO hypersurface was proposed in Ref.~\cite{bib4}:
\begin{equation}\label{CO}
k^{\ro} \frac{d^3N_i}{dk^3}~=
~\frac{d_i}{(2\pi)^3}\int_{\Sigma_f}d\sigma_{\mu}k^{\mu}~
\phi_i\left(\frac{k_{\lambda}u^{\lambda}_g -
\mu_i^g}{T_g}\right)~\theta (d\sigma_{\nu}k^{\nu})~.
\end{equation}
Eq.~(\ref{CO}) looks like CF formula (\ref{CF}), but without negative
particle numbers that appear in (\ref{CF}) for t.l. FO surfaces.
These negative contributions are cut-off by the $\theta$-function
in Eq.~(\ref{CO}). We'll call Eq.~(\ref{CO}) the cut-off (CO)
distribution in what follows.
An inclusion of the CO FO into the self-consistent
hydrodynamical scheme is considered in Ref.~\cite{bib10}.
The distribution function
for the free particles in Eq.~(\ref{CO}) contains the new parameters
$T_g, \mu_i^g, u^{\nu}_g$.
We'll briefly call the free particle state as the `gas'
in order to distinguish it from the ordinary fluid.
Note, however, that final particle system differs from the normal
fluid and gas.
It has the non-thermal CO distribution function
which is frozen-out as
all further particle rescatterings are `forbidden' in the
post-FO gas state.

The presence of the $\theta$-function in the right hand side of
Eq.~(\ref{CO}) leads to the discontinuity
between the fluid
$T_f, \mu_i^f, u^{\nu}_f$
and the gas
$T_g, \mu_i^g, u^{\nu}_g$
variables across the t.l. FO hypersurface
to satisfy the energy-momentum and charge conservations.
We call this discontinuity the FO shock.

Let us consider this FO shock on t.l. hypersurface in more detail.
To obtain the analytical solution of the problem we restrict
ourself to the case of zero baryonic number $n=0$
and consider massless particles
with the J\"uttner distribution function
\begin{equation}\label{jut}
\phi\left(\frac{k_{\mu}u^{\mu}}{T}\right)~
=~\exp\left(-~\frac{k_{\mu}u^{\mu}}{T}\right)~.
\end{equation}
The energy-momentum conservation between the fluid and free particles
along the t.l. hypersurface
acquires the form:
\begin{eqnarray}\label{conserv}
& &~ d\sigma_{\mu} T^{\mu \nu}_f \equiv
\frac{d\sigma_{\mu}}{(2\pi)^3} \int \frac{d^3k}{k^{\ro}}~k^{\mu}k^{\nu}~
\phi\left(\frac{k_{\nu}u^{\nu}_f}
{T_f}\right)~
= \nonumber \\
&=&~
d\sigma_{\mu} T^{\mu \nu}_g~\equiv
\frac{d\sigma_{\mu}}{(2\pi)^3}\int \frac{d^3k}{k^{\ro}}~k^{\mu}k^{\nu}~
\phi\left(\frac{k_{\nu}u^{\nu}_g}
{T_g}\right)~
\theta (d\sigma_{\lambda}k^{\lambda})~.
\end{eqnarray}
The energy-momentum tensor $T^{\mu \nu}_f$ in
 Eq.~(\ref{conserv}) has the form of Eq.~(\ref{tmunu})
with the functions $\varepsilon(T_f) =\varepsilon_f$ and
$p(T_f)=p_f$
given by
\begin{equation}\label{peps}
\varepsilon(T)~=~3 p(T) ~=~
\frac{1}{2\pi^2}\int_0^{\infty}k^2dk~k~
~\exp \left(- \frac{k}{T}\right) ~=~
\frac{3}{\pi ^2}~T^4~.
\end{equation}

In the rest frame of
the FO front, i.e. in
the Lorentz frame where $d\sigma_{\mu}$ becomes equal to $S_{tr}(0,dt)$, the
$T^{01}_g$ and $T^{11}_g$ components of the gas energy-momentum tensor
in Eq.~(\ref{conserv}) can be rewritten then in the form
\begin{equation}\label{tmunugas}
T^{01}_g ~=~ (\varepsilon ^*_g + p^*_g)u^0_gu^1_g~,~~~~
 T^{11}_g~=~(\varepsilon^*_g + p^*_g) u^1_gu^1_g~ + p^*_g ~.
\end{equation}
It also coincides with that of Eq.~(\ref{tmunu}),
but with effective values of $\varepsilon ^*_g$
and $p^*_g$ which are found to be equal to
\begin{equation}\label{epsef}
\varepsilon_g^* ~=~\varepsilon(T_g)~\frac{(1+v_g)^2}{4v_g}~,~~~~
p^*_g ~= ~p(T_g)~\frac{(1+v_g)^2(2-v_g)}{4}~,
\end{equation}
where $v_g$ is the velocity parameter of the gas in
the FO shock rest frame.
The functions $\varepsilon$ and $p$ in the right hand side
of Eqs.~(\ref{epsef}) are given by Eq.~(\ref{peps}).

Due to the same formal structure of $T^{01}_g , T^{11}_g$
(\ref{tmunugas}) and $T^{01}_f , T^{11}_f$ (\ref{tmunu})
one obtains the solution of Eq.~(\ref{conserv})
\begin{equation}\label{frvel}
v_{f}^{2} = \frac{(p_f - p_{g}^*)(\varepsilon_g^* + p_{f})}
{(\varepsilon_f - \varepsilon_{g}^*)(\varepsilon_{f} + p_g^*)} \; , \; \;
~ ~ v_g^{2} = \frac{(p_f-p^*_g)(\varepsilon_{f} + p^*_g)}
{(\varepsilon_f - \varepsilon^*_{g})(\varepsilon^*_g + p_{f})}~,
\end{equation}
which is similar to Eq.~(\ref{slvel}). Note, however,
that in contrast to
Eq.~(\ref{slvel})
the values
of $\varepsilon_g^*$ and
$p_{g}^*$ in Eq.~(\ref{frvel})
are not just the thermodynamical quantities, but
depend also on $v_g$.

It should be emphasized that Eq.~(\ref{tmunugas})
can also be introduced
for the distribution functions of massive and charged particles \cite{bib10}.
In the latter case the energy-momentum conservation leads to
the familiar expressions (\ref{frvel}) for the effective energy density and
pressure,
and the charge conservation
leads to the TA equation (\ref{TA}) for the effective charge
density.

We fix the FO hypersurface $\Sigma_f$ by the condition $T_g=const$.
The analytical solution
of Eq.~(\ref{frvel}) can be presented then in the form
\begin{equation}\label{vgtf}
v_g~=~\frac{9v_f^2 - 8v_f + 3}{3v_f^2 + 1}~,~~~~
R~\equiv~ \left(\frac{T_f}{T_g}\right)^4 ~=~
 \frac{4(3v_f^2 - 2v_f +1)^2 (v_f +1)}{(3v_f -1)(3v_f^2 +1)^2}~.
\end{equation}
It is instructive to compare these results with shock-wave
solution (\ref{slvel}) for the same ideal gas EoS (\ref{peps}).
We fix the temperature $T_g$ of the final state in s.l. shock wave
(\ref{slshock}) and consider $v_g$ and $T_f$ dependence on $v_f$:
\begin{equation}\label{vgtfshock}
v_g~=~\frac{1}{3v_f}~,~~~~
R~\equiv~ \left(\frac{T_f}{T_g}\right)^4 ~=~
\frac{3 (1-v_f^2)}{9v_f^2 - 1}~.
\end{equation}

Figs. 2 and 3 show the dependences of $v_g$ and $T_f$ on $v_f$
for the fixed temperature $T_g$ in the final state, both
for the FO shock
(\ref{vgtf}) and for the normal shock wave (\ref{vgtfshock}). The
kinematical restrictions on the gas velocity
give the same value of the minimal fluid velocity in both  shock
transitions,
$(v_f)_{min}= p_f/\varepsilon_f$. It equals to $1/3$
for the considered ideal gas EoS (\ref{peps}) of the fluid.
Figs.~2 and 3 indicate that at low values of the fluid velocity,
$v_f < 1/\sqrt{3}$,
the behavior of $v_f$ and $T_f/T_g$ for the FO shock
(\ref{vgtf}) and for the normal shock wave (\ref{vgtfshock})
is quite similar. The value of $v_f = 1/\sqrt{3}$ corresponds
to the speed of sound,
$c_s=1/\sqrt{3}$,
 in the system with ideal gas EoS (\ref{peps}).
According to the requirements given by
Eq.~(\ref{sound}) the shock
transitions at $v_f< c_s$ are mechanically unstable.
Mechanically stable solutions at $v_f> c_s$ for normal
shock waves and FO shocks have qualitatively different
behavior. For the stable normal shock wave one has
$v_g<v_f$ and $T_f<T_g$ (see Figs.~2, 3), i.e. only compression
normal shock wave
transitions '{\it f}~'$\rightarrow$'{\it g}~' would be stable.
For the FO shock transitions we find a
completely different behavior, $v_g>v_f$ and $T_f>T_g$,
illustrated in Figs.~2, 3.
In contrast to the normal shock waves, only
the rarefaction FO shock transitions
'{\it f}~'$\rightarrow$'{\it g}~' are stable.
This result is in agreement with an intuitive physical picture
of the FO as a rarefaction process.


\begin{figure}[htb]
\vspace*{0.3cm}
                \hspace*{-2.cm}\insertplot{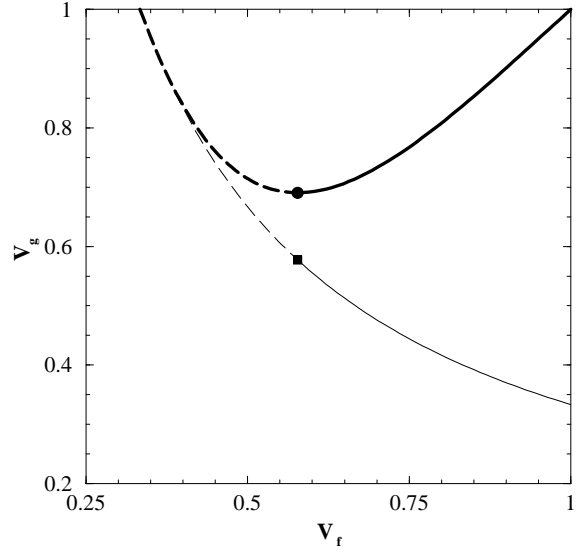}
\vspace*{-1.5cm}
\caption[]{
 Gas velocity $v_g$
as a function of
the fluid velocity $v_f$ in the rest frame of the
shock front.
The dashed lines represent mechanically unstable transitions,
whereas the solid lines show the mechanically stable FO shocks
and normal shock waves.
Thin lines correspond to a normal shock wave (Eq. (\ref{vgtfshock})) and the
  thick
ones correspond to the \fos \mbox{(Eq. (\ref{vgtf}))}.
The circle has the coordinates $(1/\sqrt{3};
3 - 4/\sqrt{3})$
and the square has the coordinates
$(1/\sqrt{3}; 1/\sqrt{3})$.
}
\label{fig2}
\end{figure}

The entropy flux of the fluid is given by
\begin{equation}\label{entropyf}
s_f^{\mu}d\sigma_{\mu}~=~s_fu_f^{\mu}d\sigma_{\mu}~,
\end{equation}
where $s_f=(\varepsilon_f+p_f)/T_f$ is the fluid
entropy density.
Similar expression is valid for the final 'gas' state
in the case of normal shock waves.
The entropy production in the normal shock waves is given by the
formula
\begin{equation}\label{entropy1}
\frac{s_g^{\mu} d\sigma_{\mu} }{s_f^{\mu} d\sigma_{\mu} }~ =~
\frac{1}{3^{3/4} v_f}
\left[ \frac{9 v_f^2 - 1}{1 - v_f^2} \right]^{1/4}\,\,
\end{equation}
and is shown in Fig.~4.
The entropy flux of the gas with the cut-off distribution
function is given by
\cite{bib10}:
\begin{equation}\label{entropyg}
s_g^{\mu}d\sigma_{\mu}~=~
\frac{d\sigma_{\mu}}{(2\pi)^3}\hspace*{-0.1cm}\int \frac{d^3k}{k^{\ro}}~k^{\mu}~
\phi\left(~\frac{k_{\nu}u^{\nu}_g}
{T_g}\right)\hspace*{-0.1cm}
\left[~1~-~\ln ~\phi \left(~\frac{k_{\nu}u^{\nu}_g}{T_g}\right)\right]
\hspace*{-0.1cm}\theta (d\sigma_{\lambda}k^{\lambda})~.
\end{equation}
The entropy production in the FO shocks is calculated then as
\begin{equation}\label{entropy2}
\frac{s_g^{\mu} d\sigma_{\mu} }{s_f^{\mu} d\sigma_{\mu} }~ =~
\frac{1}{2 v_f}
\left[ \frac{(3 v_f^2 + 1)^2(3 v_f - 1)}{(1 + v_f)} \right]^{1/4}\,\,.
\end{equation}
The maximal entropy production in the FO shock
(\ref{entropy2}) corresponds to $v_f=c_s=1/\sqrt{3}$
and can be considered as an analog of the Chapman--Jouguet
point (see Fig.~4).

\begin{figure}[htb]
\vspace*{0.1cm}
                 \insertplot{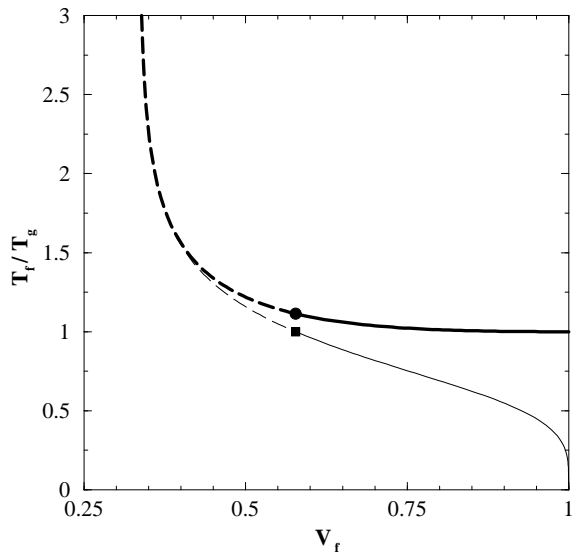}
\vspace*{-1.5cm}
\caption[]{
Ratio of the temperatures on the both
sides of the shock front, $T_f/T_g$, as a function of
the fluid velocity $v_f$ in the rest frame of the
shock front.
The legend corresponds to the Fig. 2.
The circle has the coordinates
$\left[1/\sqrt{3}; (8/(3 \sqrt{3}))^{1/4} \right]$
and the square has the coordinates $(1/\sqrt{3}; 1)$.
}
\label{fig3}
\end{figure}

\section{Conclusions}
In the present paper
the discontinuities
in relativistic hydrodynamics
have been considered.
The particle emission
from t.l. parts of the FO hypersurface
looks as a discontinuity
in the hydrodynamical motion (FO shocks). The connection
of these FO shocks to the normal shock waves
in the relativistic hydrodynamics
has been analyzed.
The above consideration can be used
in the hydrodynamical approach to the relativistic A+A collisions.
It should be also important in
the models of A+A collisions which combine hydrodynamics for the early
stage of the reaction with a microscopic
hadron description of the later
stages (see, for example, Ref.~\cite{bib20}).

\vskip 10pt
\noindent \large {\bf Acknowledgment}\normalsize \vskip 10pt
\noindent
Authors thank D.H. Rischke and G. Yen for useful discussions.
M.I.G. acknowledges the financial support of DFG, Germany.
K.A.B. is grateful to the Alexander von Humboldt Foundation for
the financial support.

\begin{figure}[htb]
\vspace*{0.1cm}
                 \insertplot{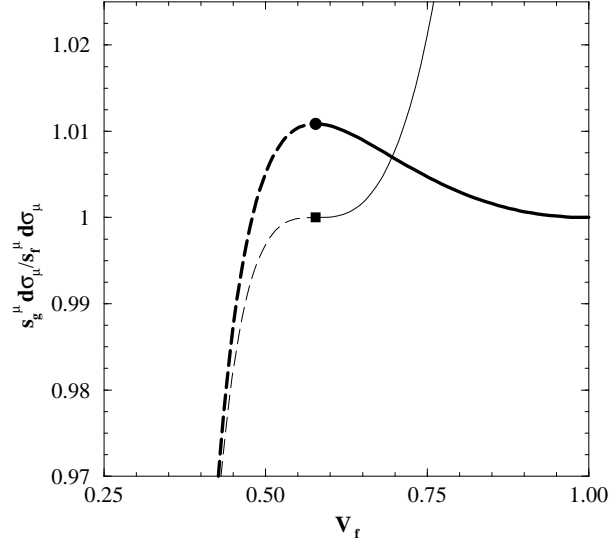}
\vspace*{-1.5cm}
\caption[]{
Ratio of the entropies on the both
sides of the shock front, as a function of
the fluid velocity $v_f$ in the rest frame of the
shock front.
The legend corresponds to the Fig. 2.
The circle corresponds to the maximal entropy
(analog of the Chapman-Jouguet point)
in the
\fos and it has the coordinates
$\left[1/\sqrt{3};
\left(\frac {9 (3 - \sqrt{3}) }{4 (\sqrt{3} + 1)}\right)^{1/4} \right]$
and the square has the coordinates $(1/\sqrt{3}; 1)$.
}
\label{fig4}
\end{figure}

\vfill\eject
\end{document}